\documentclass[twocolumn,showpacs,preprintnumbers]{revtex4}
\usepackage{graphicx}
\usepackage{dcolumn}
\usepackage{bm}

\begin{document}

\title{New manifestation of the dispersion relation: Breakup Threshold
Anomaly.}
\author{M. S. Hussein$^{1}$, P. R. S. Gomes$^{2}$, J. Lubian$^{2}$ and 
L. C. Chamon$^{1}$.}
\affiliation{$^1$Instituto de F\'{\i}sica da Universidade de S\~{a}o Paulo, 
Caixa Postal 66318, 05315-970, S\~{a}o Paulo, SP, Brazil. \\
$^2$Instituto de F\'{\i}sica, Universidade Federal Fluminense, Av. Litoranea
s/n, Gragoat\'a, Niter\'oi, RJ, Brazil, 24210-340.}
\date{\today }

\begin{abstract}
It is pointed out that the usual threshold anomaly, found operative in the
energy behavior of the imaginary and real parts of the optical potential
representing the elastic scattering of tightly bound nuclei at near- and
below-barrier energies, suffers a drastic qualitative change in the case of
the elastic scattering of weakly bound nuclei. Owing to the strong coupling
to the breakup channel even at sub-barrier energies, the imaginary potential
strength seems to increase as the energy is lowered down to below the
natural, barrier, threshold, accompanied by a decrease in the real potential
strength. This feature is consistent with the dispersion relation. The
system $^{6}$Li + $^{208}$Pb is analyzed to illustrate this new phenomenon.
\end{abstract}

\maketitle

The, by now, well known Threshold Anomaly (TA), seen in the behavior of the
real and imaginary parts of the optical potential as a function of
decreasing energy in the elastic scattering of tightly bound nuclei at
near-barrier energies, has been discussed and reviewed by several authors 
\cite{0,2,1}. The phenomenon is a direct consequence of the dispersion
relation which quantifies the concept of causality in scattering: no
scattered wave emerges before the incident wave reaches the target.
Recently, the TA has been looked for in the case of the elastic scattering
of weakly bound stable and radioactive nuclei \cite{3,3b,3c,3d,3e,3f,3g}.
Careful analyses of the data show that what happens in these systems is a
new manifestation of the dispersion relation unique for breakup coupling
dynamic polarization potential. Because the coupling to the breakup in these
systems continues to be important even at energies below the barrier, the
"threshold" ceases to be the barrier itself. Thus, the imaginary part of the
potential could increase at lower energies and, as the dispersion relation
dictates, the real part of the dynamic potential would show a decrease,
implying an overall decrease in the real part of the optical potential that
fits the elastic scattering. This is indeed what is found 
\cite{3,3b,3c,3d,3e,3f,3g}. The purpose of this paper is to give an account of
this phenomenon, which we coin as the Breakup Threshold Anomaly (BTA).

We analyze a system involving the $^{6}$Li weakly bound nucleus: $^{6}$Li + 
$^{208}$Pb. We have re-analyzed the elastic scattering angular distribution 
data \cite{3c} using the S\~{a}o Paulo (SP) optical potential. The choice of 
this potential is due to the fact that it has been successfully used in the 
study of several systems \cite{10}. The conclusions obtained here, however, do 
not depend on this particular choice of potential. The SP interaction 
\cite{10,7,8,9,11} is based on the double folding potential with account to 
exchange through an effective energy dependence: 
\begin{equation}
V_{SP}(R,E)=\left( 1+i\;0.78\right) F(R,E),
\end{equation}%
where $F(R,E)$ is the double folding potential whose energy dependence
results from the local equivalence of the otherwise non-local interaction 
\cite{11,12}. This energy dependence is not dispersive. The $F(R,E)$ term is
given by:
\begin{equation}  
F(R,E) = V_F(R) \; e^{- 4 \mbox{v}^2/c^2} \;,  
\end{equation}  
where $c$ is the speed of light and $\mbox{v}$ is the local relative velocity 
between the two nuclei,  
\begin{equation}  
\mbox{v}^2(R,E) = \frac{2}{\mu} \, 
\left[ E-V_C(R)-V_{N}(R,E) \right] \;, 
\end{equation}  
where  $V_N$ is the real part of the nuclear interaction and $V_C$ is the 
Coulomb potential. The folding potential depends on the matter densities of the 
nuclei involved in the collision:
\begin{equation}
V_F(R) = \int \rho_1(\vec{r}_1) \; \rho_2(\vec{r}_2) \; 
V_{0} \; \delta(\vec{R}-\vec{r_{1}}+\vec{r_{2}}) \; d\vec{r_1} , 
\end{equation}
with $V_0 = - 456$ MeV fm$^3$. The use of the matter densities and delta 
function in Eq. (4) corresponds to the zero-range approach for the folding 
potential, which is equivalent \cite{9} to the more usual procedure of using 
the M3Y effective nucleon-nucleon interaction with the nucleon densities of the 
nuclei (instead of the matter densities). With the aim of providing a parameter 
free description of the interaction, we proposed \cite{9} an extensive 
systematics of nuclear densities. For this purpose, we adopted the two-parameter 
Fermi (2pF) distribution to describe the densities. The radii of the 2pF 
distributions are well represented by
\begin{equation}
R_0=1.31 A^{1/3}-0.84 \; \mbox{fm} ,
\end{equation}
where $A$ is the number of nucleons of the nucleus. The values obtained for the 
matter diffuseness of the distributions are very similar throughout the 
periodic table and present small deviations around the average value 
$a = 0.56$ fm. In this work, we use the S\~ao Paulo potential in the 
context of this systematics by assuming the average diffuseness value and Eq. 
(5) to determine the radii of the density distributions.

In the present analysis, we have assumed for the optical potential a normalized 
version of the SP potential: 
\begin{equation}
V_{SP}(R,E)=\left[ N_{R}(E)+iN_{I}(E)\right] F(R,E).
\end{equation}%
The coefficients $N_{R}(E)$ and $N_{I}(E)$ are energy dependent
normalization factors that take into account the effects of the dynamic
polarization potentials (DPP) arising from direct channel couplings. It is
worth mentioning here that all DPP%
\'{}%
s are dispersive with their real and imaginary parts being connected through
a dispersion relation. One important exception to this is the elastic
transfer DPP \cite{13}.

From the properties of the Green function, that enters in the definition of
the DPP, one can immediately derive the dispersion relation between $N_{R}(E)$ 
and $N_{I}(E)$, represented by: 
\begin{equation}
N_{R}(E)=N_{R0}+\Delta N_{R}(E),
\end{equation}%
\begin{equation}
\Delta N_{R}(E)=\frac{P}{\pi }\int \frac{N_{I}(E^{\prime })}{E^{\prime }-E}%
dE^{\prime },
\end{equation}%
and its subtracted form 
\begin{equation}
\Delta N_{R}(E)=\Delta N_{R}(E_{s})+(E-E_{s})\frac{P}{\pi }\int \frac{%
N_{I}(E^{\prime })}{(E^{\prime }-E_{s})(E^{\prime }-E)}dE^{\prime },
\end{equation}%
where $E_{s}$ is some high enough energy at which information about both $%
N_{I}$ and $\Delta N_{R}$ are known \cite{2}. These equations are the analog
of the Kramer-Kronig dispersion relation in optics, from general principle
of causality, already mentioned above.

The dashed lines in figure 1 represent predictions, for the elastic scattering
of $^{6}$Li + $^{208}$Pb, obtained with the energy-independent standard values 
$N_{R}$ = 1 and $N_{I}$ = 0.78 , while the solid lines correspond to the results
obtained with the best fit $N_{R}$ and $N_{I}$ values. A comparison between
these dashed and solid lines shows that the data fit is quite sensitive to
the $N_{R}$ and $N_{I}$ values.

Just to illustrate the dispersion relation, we have assumed a schematic 
description for $N_I$
\begin{equation}
N_I=0 \; \mbox{for} \; E \le E_1 ,
\end{equation}
\begin{equation}
N_I= a (E-E_1) \; \mbox{for} \;E_1 \le  E \le E_2 ,
\end{equation}
\begin{equation}
N_I= a (E_2-E_1) + b (E-E_2) \; \mbox{for} \; E_2 \le  E \le E_3 ,
\end{equation}
\begin{equation}
N_I= a (E_2-E_1) + b (E_3-E_2) = N_{I\infty} \; \mbox{for} \; E \ge E_3 .
\end{equation}
Within this assumption, and owing to the constancy of $N_I$ at $E>E_3$, the
subtracted dispersion relation, Eq. (9), gives the same result as the
non-subtracted one, Eq. (8). Using Eq. (8) and Eqs. (10-13), one obtains an 
analytical expression for $\Delta N_R$ 
\cite{2,1}
\begin{eqnarray}
\Delta N_R(E) &=& a (E_2-E_1) \left[ \epsilon_1 ln|\epsilon_1| -
\epsilon_2 ln| \epsilon_2| \right] \\ \nonumber
& & + \left[ b (E_3-E_2) -  a (E_2-E_1) \right] \\ \nonumber
& & \times \left[ \epsilon'_2 ln|\epsilon'_2| -
\epsilon'_3 ln| \epsilon'_3| \right] ,
\end{eqnarray}
with $\epsilon_i=(E-E_i)/(E_2-E_1)$ and $\epsilon'_i=(E-E_i)/(E_3-E_2)$. By
using Eqs. (7) and (14), one can find $N_{R}(E)$ in an independent way of
the reference energy $E_S$.

In figure 2 we present the best fit $N_{R}(E)$ and $N_{I}(E)$ values. The
uncertainties of these quantities have been obtained by considering the range 
where $N_{R}(E)$ and $N_{I}(E)$ could vary which would result in an increase 
of the chi-square value by unity relative to the corresponding minimum value.
The lines in figure 2 represent possible behaviors of $N_{R}$ and $N_{I}$ that 
are compatible with the dispersion relation. A striking difference in the 
energy dependence of these normalization coefficients from the usual energy 
dependence of tightly bound systems is clearly seen. As the energy is lowered 
below the barrier $N_{R}(E)$ decreases, while $N_{I}(E)$ increases. This 
implies an effective reduction of the nuclear attraction leading to an 
increase in the barrier height.

A further evidence of the consistency of our analyses are the values
obtained for $N_{R0}$ and $N_{I\infty }$. As the SP potential, equation (1),
has been successful in describing the elastic scattering for a large number
of different systems at energies above the barrier \cite{10}, one should
expect that both $N_{R0}$ and $N_{I\infty }$ to be close to unity. Of
course, the values for $N_{R0}$ and $N_{I\infty }$ found in the present work
are not identical to the standard (from high energies) $N_{R}$=1 and $N_{I}$%
=0.78. However, these standard values in fact represent mean values obtained
from data analyses of several systems \cite{10} and one can expect
variations around the average values for particular systems. Indeed,
structure effects on the nuclear densities involved in the folding
calculations may affect $N_{R}$ while different degrees of absorption from
particular reaction channels could affect $N_{I}$. An inspection of Fig. 2 
shows that $N_{R0} \approx 1$ and $N_{I\infty } \approx 1$ have been 
found in the present work.

As already commented, the solid lines in Fig. 2 represent behaviors
of $N_{R}$ and $N_{I}$ compatibles with the dispersion relation. In fact,
different behaviors, that also could follow the "data", can be found.
Therefore, clearly our findings are mainly based on the $N_{R}$ and $N_{I}$
"data" themselves. The detected difference between the normal and
weakly-bound systems is mostly based on the results from the corresponding
lowest energies (see Fig. 2). Even so, we consider that significant
evidence for the proposed BTA has been obtained, because the data fits for
these low energies are quite sensitive to $N_{R}$ and $N_{I}$, as
illustrated by the dashed and solid lines in Fig. 1. We mention that similar
behavior of the increasing of the imaginary potential as the energy
decreases towards the barrier was already observed earlier for the same
system \cite{3f} using different data analysis procedures and also for the
following other systems: $^{9}$Be + $^{209}$Bi \cite{3,3b}, $^{6}$Li + $%
^{28} $Si, $^{58}$Ni, $^{122}$Sn, $^{138}$Ba \cite{3f,3d} and $^{9}$Be + 
$^{64}$Zn \cite{3g}, although in these previous works there is no attempt to 
explain the behavior of the energy dependence of the real and imaginary parts 
of the optical potential as the BTA phenomenon.

Traditionally, the threshold anomaly is formally displayed in terms of
complex renormalization factors of the double-folding potential, as we have
done in the present work. This assumption, which corresponds to assuming that
the radial dependence of the optical potential is the same as the bare one,
is well established for normal tightly bound stable nucleus systems. The
situation is more complicated in the case of weakly bound nuclei, where the
effect of the couplings give rise to a polarization potential with different
radial shape. In particular, an important role is played by the tail of the
polarization which is much longer than that of the folding potential \cite%
{reff}. Nevertheless, our simple approach of considering renormalization
factors has provided good fits for all angular distributions analyzed in the
present work

In conclusion, we have found some evidence of differences in the
energy-dependence of the optical potential for tightly and weakly bound
nuclei. The Break-up Threshold Anomaly implies an increase of the imaginary
part and a decrease of the real part of the optical potential that fits the
elastic scattering at low energies. Our findings have mostly been based on
the results obtained from elastic scattering data analyses for the lowest
energies of one system, and also on earlier data analyses for other similar
systems \cite{3,3b,3c,3d,3e,3f,3g}. Clearly more work is required to further
improve our understanding of the BTA.

\begin{acknowledgments}
This work was partially supported by Financiadora de Estudos e Projetos
(FINEP), Funda\c{c}\~{a}o de Amparo \`a Pesquisa do Estado de S\~ao Paulo
(FAPESP), Funda\c{c}\~{a}o de Amparo \`a Pesquisa do Estado do Rio de
Janeiro (FAPERJ), Instituto de Mil\^{e}nio de Informa\c{c}\~{a}o Qu\^{a}%
ntica (MCT), and Conselho Nacional de Desenvolvimento Cient\'{\i}fico e
Tecnol\'ogico (CNPq).
\end{acknowledgments}

\newpage

\begin{figure}[tbp]
\vspace{-1.5cm} \hspace{-13.0cm} \includegraphics{NFIG1.ps} \vspace{8.0cm}
\caption{Elastic scattering angular distributions for the $^{6}$Li + $^{208}$%
Pb system (data from \cite{3c}). The solid lines represent the optical model 
results obtained considering the best fit $N_{R}$ and $N_{I}$ parameters. The 
dashed lines correspond to the SP potential with the standard $N_{R}=1$ and 
$N_{I}=0.78$ values.}
\end{figure}

\begin{figure}[tbp]
\vspace{10.5cm} \hspace{-13.0cm} \includegraphics{NFIG2.ps} \vspace{-2.5cm}
\caption{Energy dependence of the normalization factors $N_{R}$ and $N_{I}$
of the SP potential for the $^{6}$Li + $^{208}$Pb system. The lines
represent possible behaviors of $N_{R}$ and $N_{I}$ that are compatible with
the dispersion relation.}
\end{figure}

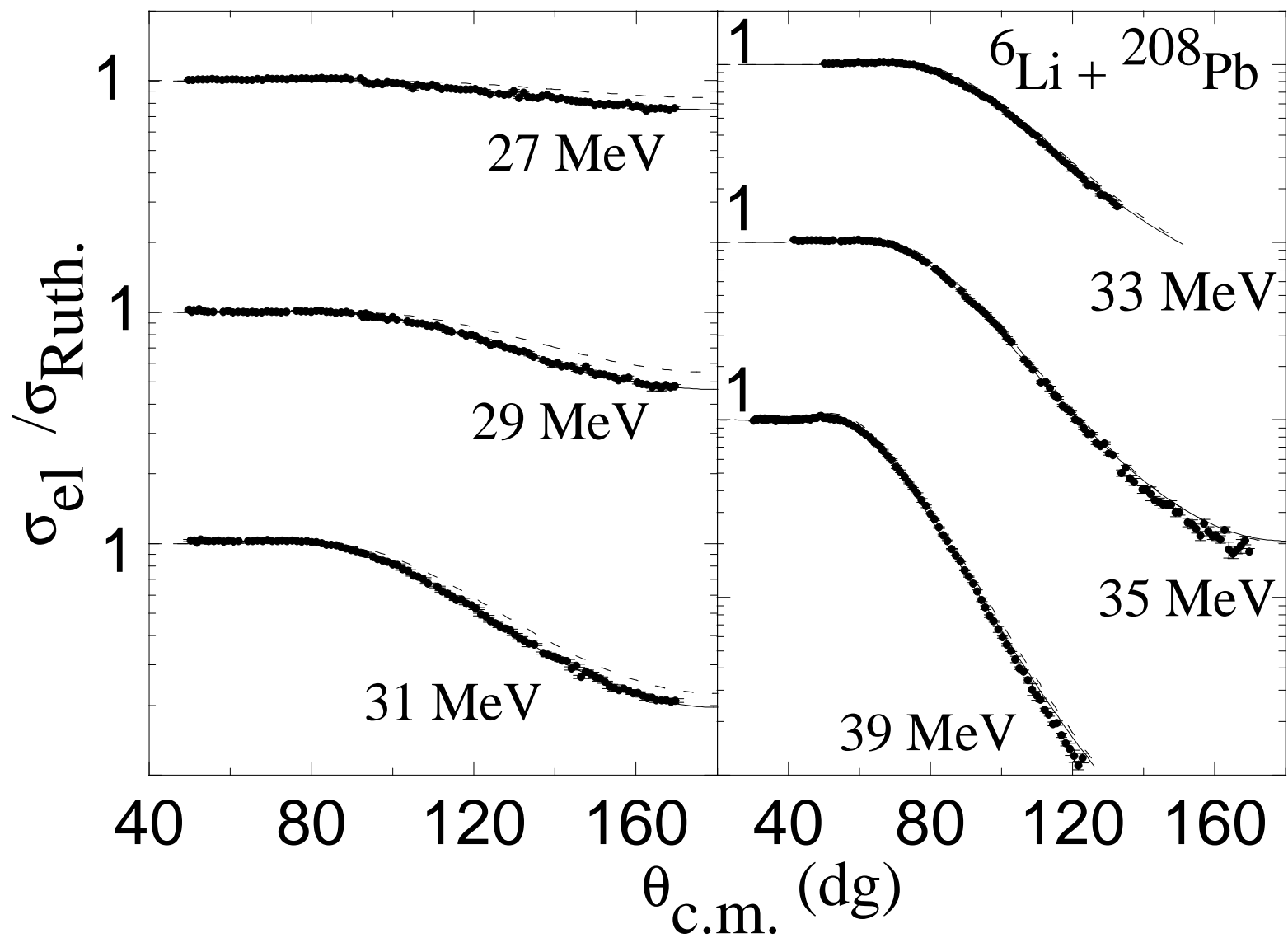
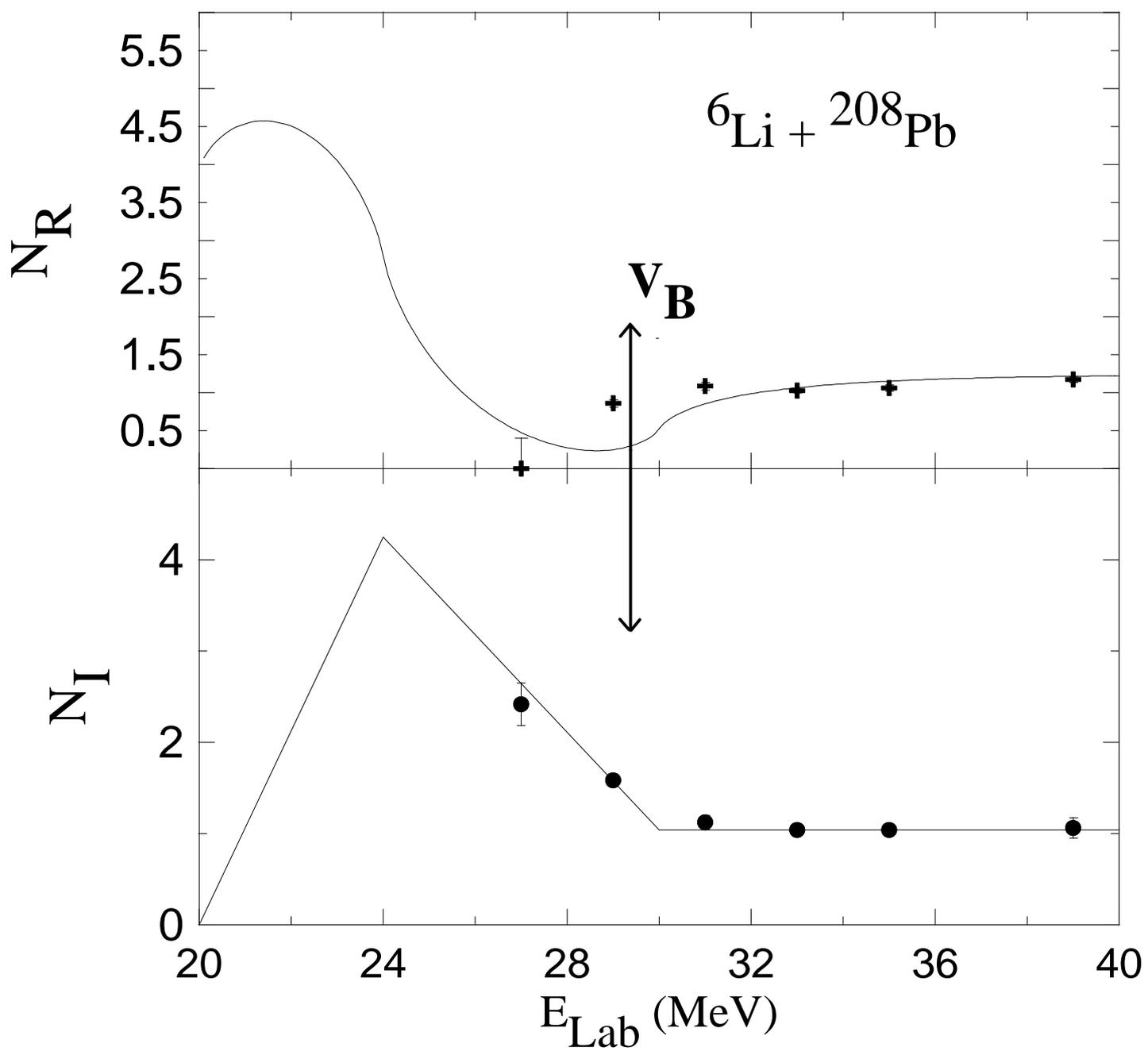
\end{document}